\documentclass{ws-procs9x6}

\begin{document}

\title{
Photoabsorption and Photoproduction on Nuclei in the Resonance Region
}

\author{
S. Schadmand
}

\address{
Institut f\"ur Kernhysik \\
Forschungszentrum J\"ulich \\
D-52425 J\"ulich, Germany \\
E-mail: s.schadmand@fz-juelich.de
}

\maketitle

\abstracts{
Inclusive studies of nuclear photoabsorption have provided
clear evidence of medium modifications in the properties of hadrons.
However, the results have not been explained in a model independent way.
A deeper understanding of the situation is expected from a detailed
comparison of meson photoproduction from nucleons and from nuclei
in exclusive reactions. Recent experimental results are presented.
}

\section{Introduction}

Current issues in the understanding of the strong interaction
address the structure of hadrons, consisting of quarks and gluons,
as the building blocks of matter.
Central challenges concern the questions why quarks are confined
within hadrons and how hadrons are constructed from their constituents.
One goal is to find the connection between the parton degrees of freedom
and the low energy structure of hadrons leading to the study of the
hadron excitation spectrum and the search for exotic
states, like glueballs or hybrid states.
An approach related to the question of the origin of hadron masses
is the search for modifications of hadron properties in the nuclear medium.
The underlying question is the origin of hadron masses
in the context of chiral symmetry breaking.
Evidence for such effects has been searched for in many experiments.
In this contribution, photoabsorption and meson photoproduction on nuclei
are discussed.

\section{Nuclear Photoabsorption}

\begin{figure}[htb]
\centerline{\epsfxsize=0.45\textwidth\epsfbox{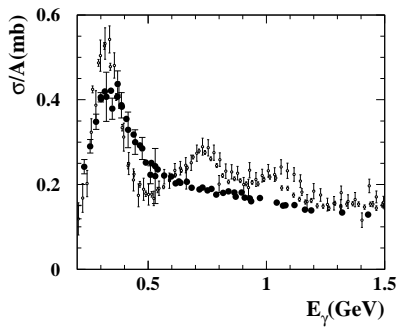}}
\caption{
Nuclear photoabsorption cross section per nucleon
as an average over the nuclear systematics
\protect\cite{Muccifora:1998ct} (full symbols)
compared to the absorption on the proton
\protect\cite{Hagiwara:2002fs} (open symbols).
}\label{fig:photoabs-nucs}
\end{figure}

Photoabsorption experiments on the free nucleon demonstrate
the complex structure of the nucleon and its excitation spectrum,
as shown in Fig.~\ref{fig:photoabs-nucs}.
The lowest-lying peak is the $\Delta$(1232) resonance
and is prominently excited by incident photons of 0.2--0.5~GeV.
The following group of resonances, P$_{11}$(1440), D$_{13}$(1520),
and S$_{11}$(1535), is called the second resonance region
(E$_\gamma$=0.5-0.9~GeV), a third resonance region is visible.
The observed resonance structures
have been studied using their decay via light mesons, showing that
the photoabsorption spectrum can be explained by the sum of $\pi$, $\pi\pi$ and
$\eta$ production cross sections.

Fig.~\ref{fig:photoabs-nucs} also
shows the nuclear photoabsorption cross section per nucleon
as an average over the nuclear systematics~\cite{Muccifora:1998ct}.
The $\Delta$ resonance is broadened and slightly shifted while
the second and higher resonance regions seem to have disappeared.

Mosel et al.~\cite{Mosel:1998rh,Lehr:1999zr,Effenberger:1997rc}
have argued that an in-medium broadening of the D$_{13}$(1520)
resonance is a likely cause of the suppressed photoabsorption
cross section.
Hirata et al.~\cite{Hirata:2001sw} see a change of the interference
effects in the nuclear medium as one of the most important reasons
for the suppression of the resonance structure.
However, the absence of resonance structure in nuclear
photoabsorption has not been explained in a model-independent way.
A deeper understanding of the situation is anticipated from the
experimental study of meson photoproduction on
nucleons embedded in nuclei in comparison to studies on the free nucleon.

\section{Meson Production in the Second Resonance Region}

In the second resonance region, double pion production aims at
the resonances D$_{13}$(1520) and P$_{11}$(1440) while
$\eta$ production is characteristic for the S$_{11}$(1535) resonance.
The three resonances in the second resonance
region decay to roughly 50\% via single pion emission.
The most trivial medium modification is the broadening of the excitation
functions due to Fermi motion.
The decay of the resonances is further modified by Pauli-blocking
of final states, which reduces the resonance widths.
In addition, decay channels like $\mbox{N}^{\star}\mbox{N}\rightarrow \mbox{NN}$
cause collisional broadening.
Both effects could cancel to some extent.

\begin{figure}[hbt]
\epsfxsize=0.48\textwidth\epsfbox{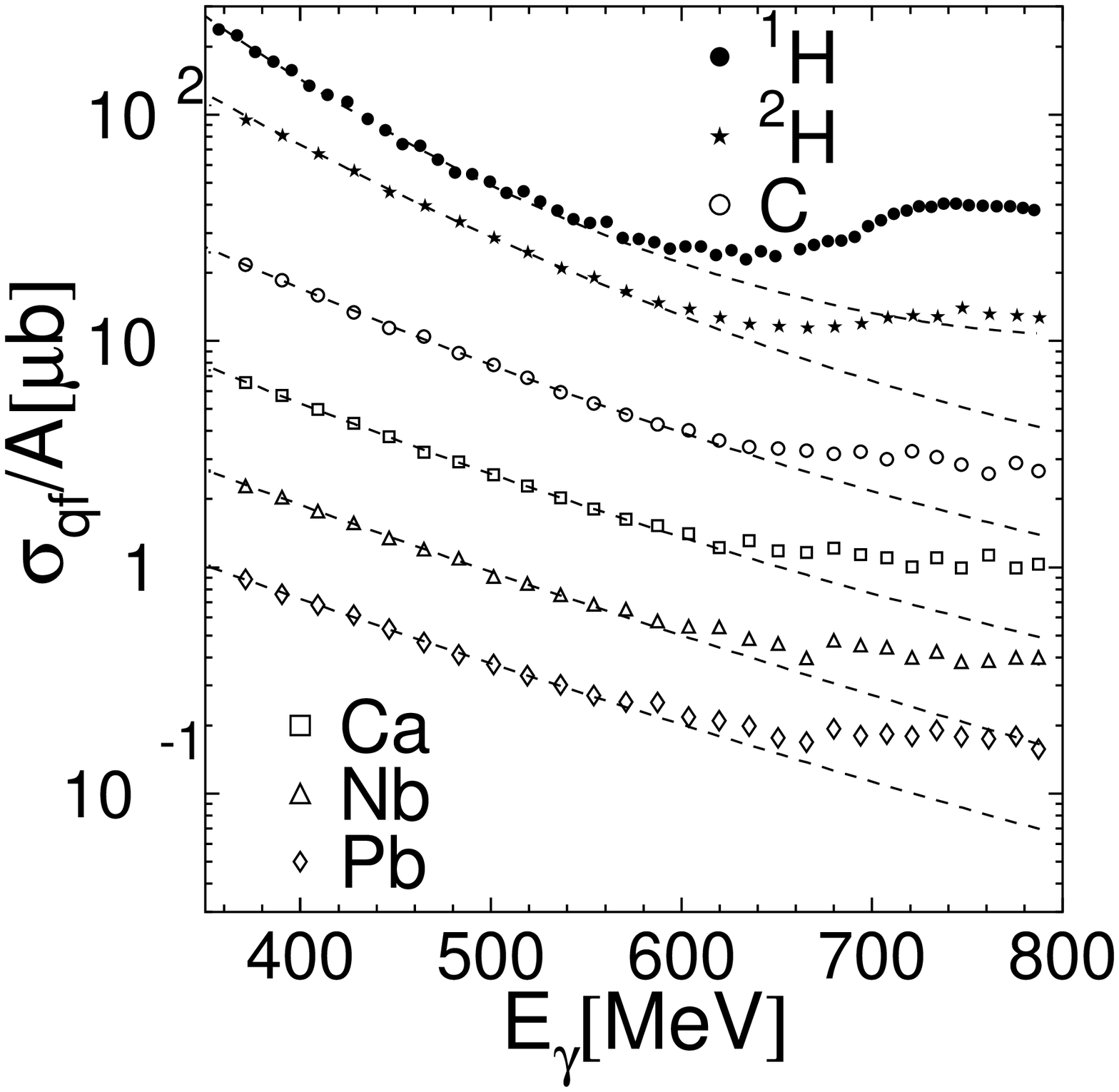}
\epsfxsize=0.51\textwidth\epsfbox{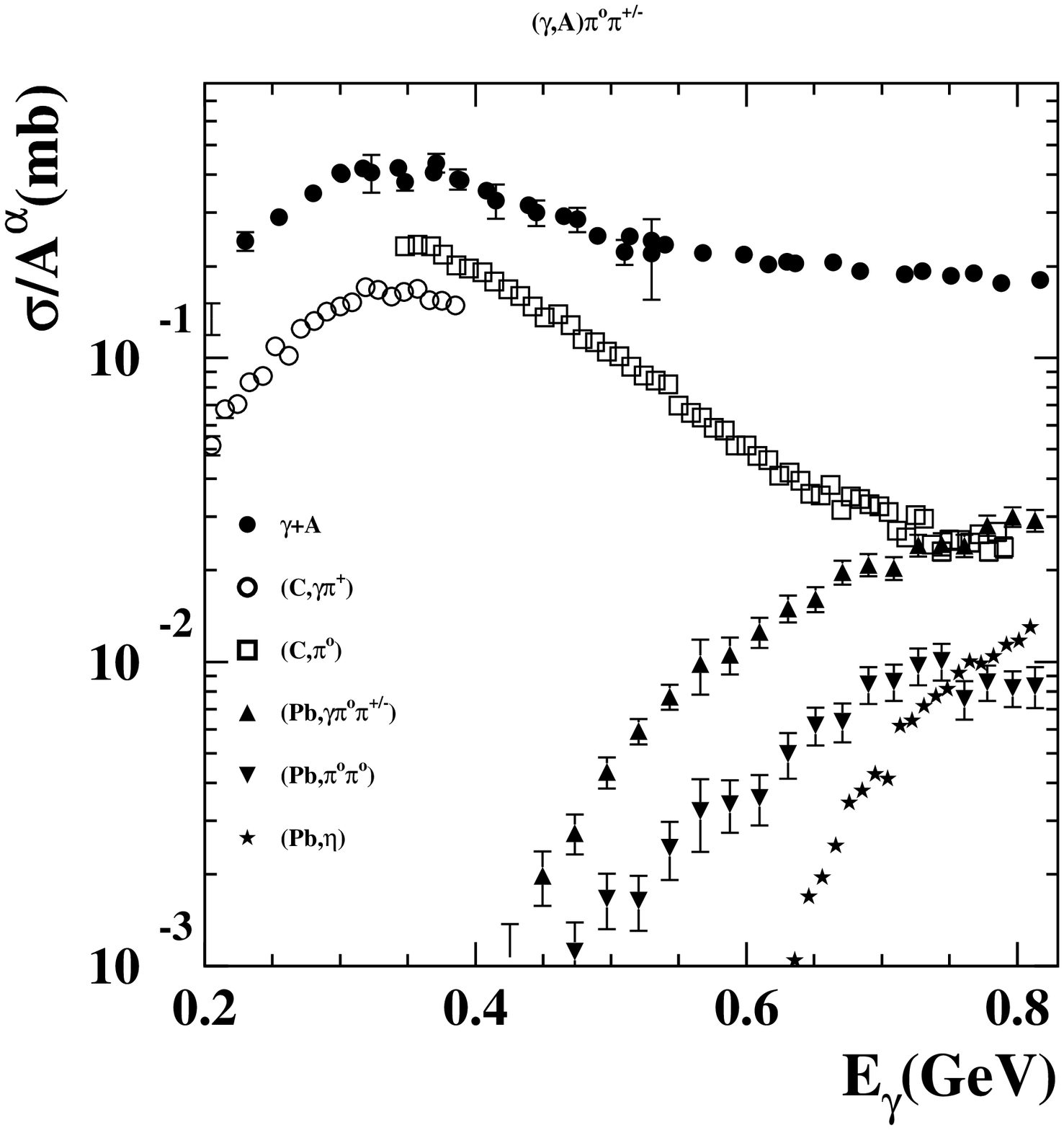}
\caption{
Left:
Total cross section per nucleon for single $\pi^\circ$
photoproduction in the second resonance region for the nucleon and for nuclei.
The scale corresponds to the proton data, the other data are scaled down
by factors 2,4,8,16,32, respectively.
The dashed curves are fits to the data in the energy range 350--550~MeV.
Right:
Status of the decomposition of nuclear photoabsorption into
meson production channels (scaled with A$^\alpha$, $\alpha$=2/3).
Small open circles are the average nuclear photoabsorption cross section
per nucleon ($\alpha$=1) \protect\cite{Muccifora:1998ct}.
Meson production data are from
\protect\cite{Arends:1982ed,Krusche:2001ku,Roebig_Landau:1996xa,Yamazaki:2000jz,Janssen:2002,Schadmand:2005}.
}\label{fig:exp-nucs}
\end{figure}
An attempt to study the in-medium properties of
the D$_{13}$ resonance was undertaken with a measurement of quasifree single
$\pi^\circ$ photoproduction \cite{Krusche:2001ku} which, on the free nucleon,
is almost exclusively sensitive to the D$_{13}$ resonance.
The left panel of Fig.~\ref{fig:exp-nucs} summarizes the results.
Strong quenching of the D$_{13}$-resonance structure
is found for the deuteron with respect to the nucleon.
However, an indication of a broadening or a suppression
of the D$_{13}$ structure in heavy nuclei is not observed.
Model predictions agree with the pion photoproduction
data only under the assumption of a strong broadening of the resonance,
other effects seem to be missing in the models. This casts doubt
on the interpretation of the total photoabsorption data via resonance
broadening.
In contrast to the case of total photoabsorption, the second resonance
bump remains visible.
However, exclusive reaction channels are dominated by the nuclear
surface region where in-medium effects are smaller.
Furthermore, as discussed in \cite{Lehr:2001ju},
resonance broadening effects are even more diluted for reactions
which do not contribute to the broadening,
due to the averaging over the nuclear volume.

The right panel of Fig.~\ref{fig:exp-nucs} shows the status
of the decomposition of nuclear photoabsorption into meson production
channels.
The available experimental meson cross sections are exclusive
measurements, investigating quasifree production.
The purely charged final states have not been measured.
However, it can be inferred from the existing data
that the sum of the cross sections would not reproduce
the flat shape of the total photoabsorption from nuclei.
In a recent compilation \cite{Krusche:2004zc}, it is observed that
the current results indicate large differences between quasi\-free meson
production from the nuclear surface and non-quasifree components.
The quasifree part does not show a suppression of the resonance structures
in the second resonance region.
However, resonance structures seem absent in the
non-quasifree meson production which has larger contributions from the
nuclear volume.

\section{$\omega$ Mesons in the nuclear Medium}

The photoproduction of $\omega$ mesons on nuclei has been
investigated using the Crystal Barrel/TAPS experiment
at the ELSA tagged photon facility in Bonn \cite{Trnka:2005ey}.
The aim is to study possible
in-medium modifications of the $\omega$ meson via the reaction
$\gamma + A \rightarrow \omega + X \rightarrow \pi^\circ \gamma + X^\prime$.
A number of theoretical models predict a mass shift of the omega meson 
in the nuclear medium, for references see 
\cite{Messchendorp:2001pa,Trnka:2005ey}.
An excellent recent review discusses nucleon and 
hadron structure changes in the nuclear medium
within the quark-meson coupling (QMC)
model predicting a reduction in the omega mass in
nuclei \cite{Saito:2005rv}.
Experimentally, 
results obtained for Nb are compared to a reference measurement
on a $\rm{LH_2}$ target.
\begin{figure}[htb]
\epsfxsize=0.32\textwidth\epsfbox{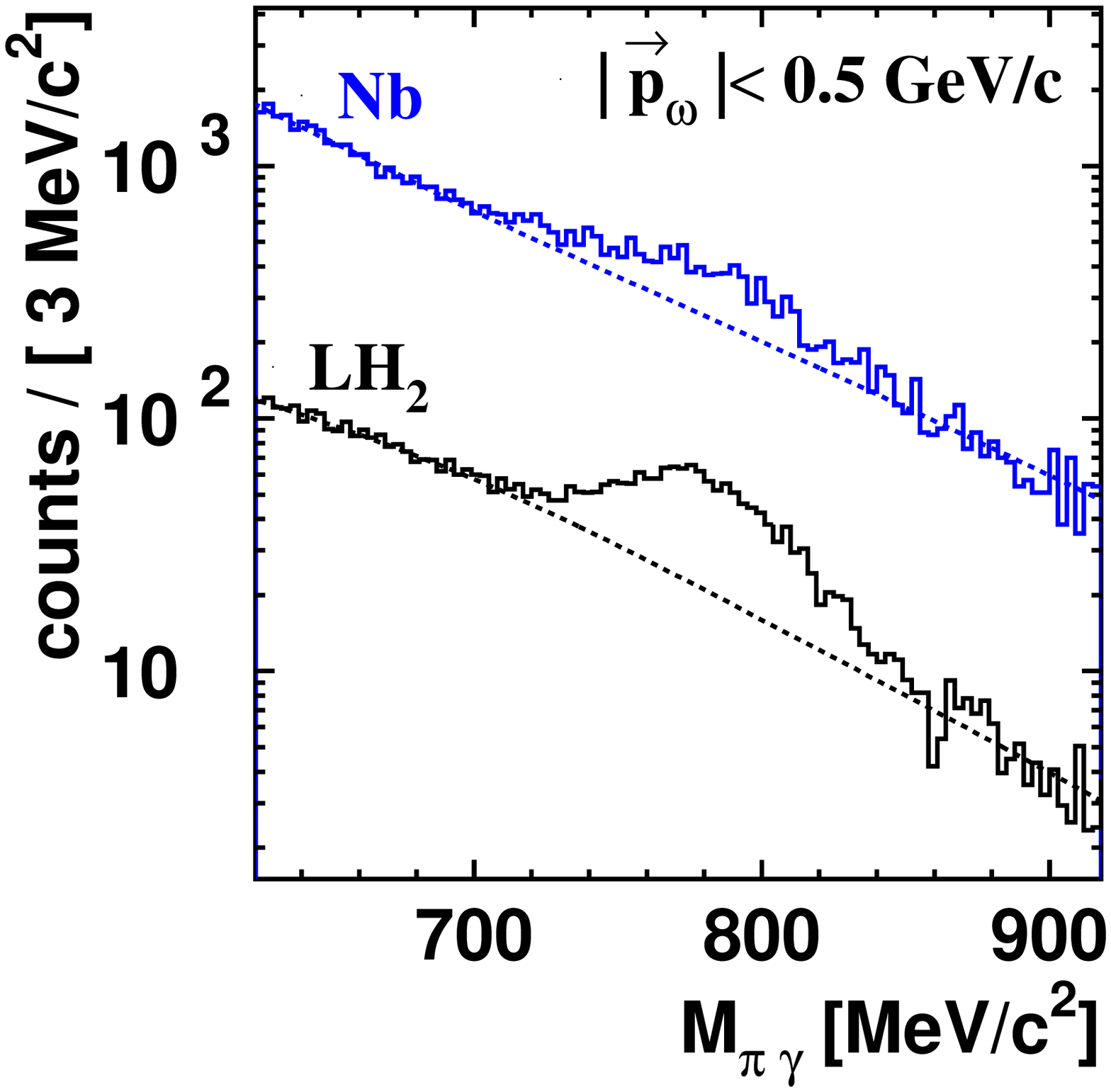}
\epsfxsize=0.32\textwidth\epsfbox{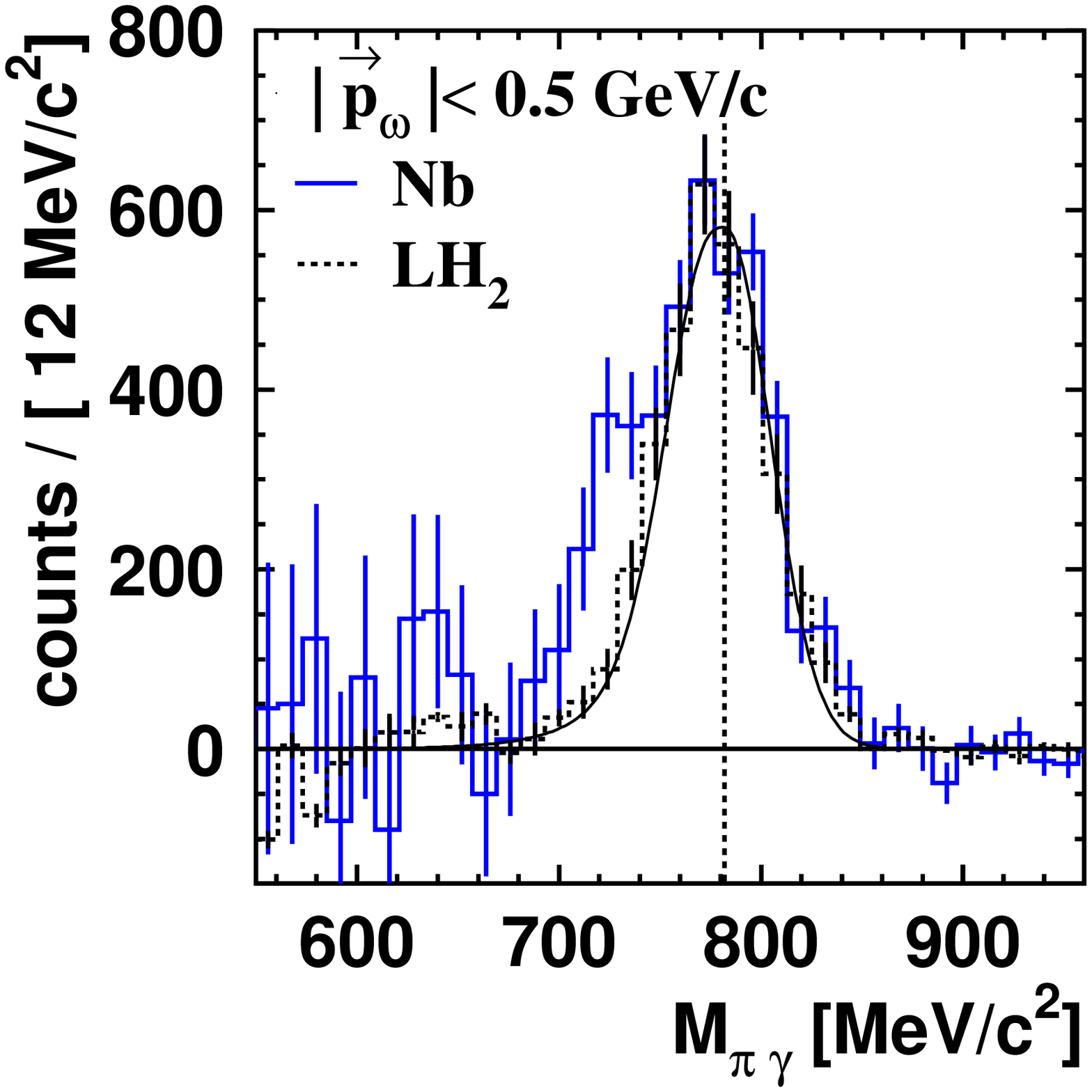}
\epsfxsize=0.32\textwidth\epsfbox{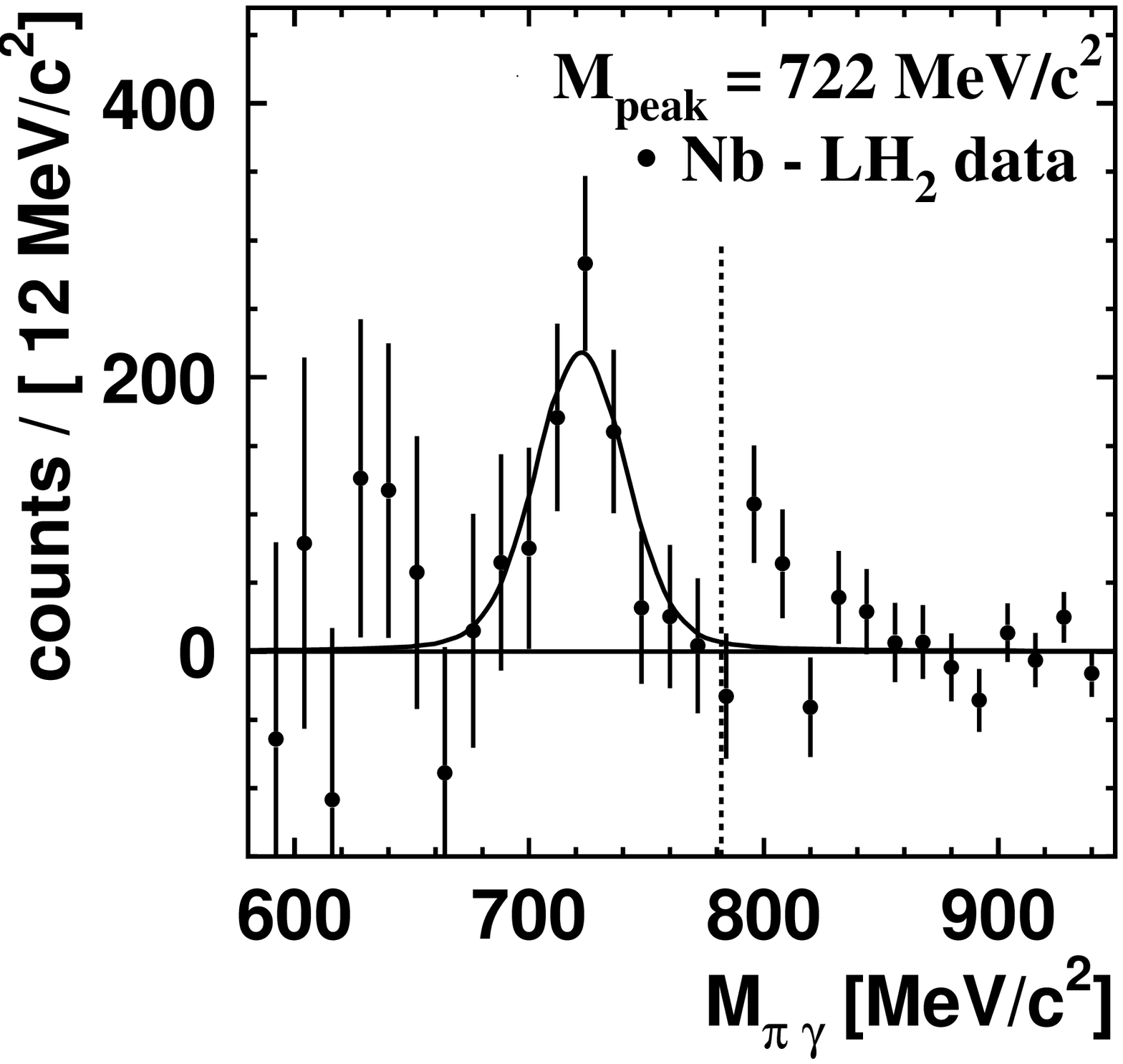}
\caption{
Left panel:
Inclusive $\pi^\circ \gamma$ invariant mass
spectra for $\omega$ momenta less than 500 $\rm{MeV/c}$.
Upper histogram: Nb data, lower histogram: $\rm{LH_2}$ target
reference measurement.
The dashed lines indicate fits to the respective background.
Middle panel:
$\pi^\circ \gamma$ invariant mass for the Nb data
(solid histogram)
and $\rm{LH_2}$ data (dashed histogram) after background subtraction.
The error bars show
statistical uncertainties only. The solid curve represents
the simulated line shape for the $\rm{LH_2}$ target.
Right panel:
In-medium decays of $\omega$ mesons along with a Voigt fit to the
data.
The vertical line indicates the vacuum $\omega$ mass of 782 $\rm{MeV/c^2}$.
}\label{fig:omega}
\end{figure}
The left panel of Fig.~\ref{fig:omega} shows the
$\pi^\circ\gamma$ invariant mass distribution
without further cuts except for a three momentum cutoff of
$|\vec{p}_{\omega}| < \rm{500~MeV/c}$.
The dominant background source is two pion production where one
of the four photons escapes the detection.
This probability was determined by Monte Carlo simulations
to be $14 \%$.
The resulting three photon final state is not distinguishable from the
$\omega \rightarrow \pi^\circ \gamma$ invariant mass.
These decays are eliminated by matching the right hand part
of the Nb invariant mass spectrum to the $\rm{LH_2}$ data
(see central panel of Fig.~\ref{fig:omega})
and by subtracting the two spectra from each other.
For this normalization the integral of the undistorted spectrum
corresponds to $75 \%$ of the counts in the Nb spectrum.
This is in good agreement with a theoretical prediction obtained
from a transport code calculation \cite{Muhlich:2003tj,Muhlich:2004}.
There, about $16 \%$ of the total decays are predicted to occur inside
the nuclear medium ($\rho > 0.1\cdot \rho_0$) without
final state interaction (FSI) and $3 \%$
of the events are distorted due to FSI in the mass range of
$0.6 ~\rm{GeV/c^2} < M_{\pi^\circ \gamma} < 0.9~\rm{GeV/c^2}$.
In addition, $9 \%$ of the events are moved towards lower masses
due to the $\Delta$ decay kinematics.
The right panel of Fig.~\ref{fig:omega}
shows the invariant mass distribution obtained after
background subtraction.
The expected superposition of decays outside of the nucleus is observed
at the nominal vacuum mass with decays occurring inside the nucleus,
responsible for the shoulder towards lower invariant masses.
The high mass part of the $\omega$ mass signal appears to be identical
for the Nb and $\rm{LH_2}$ targets, indicating that this part is
dominated by $\omega$ meson decays in vacuum.

A difference in the line shape for the two data samples is not observed
for recoiling, long-lived mesons ($\pi^\circ$, $\eta$ and $\eta^\prime$),
which decay outside of the nucleus.
However, for $\omega$ mesons produced on the Nb target
a significant enhancement towards lower masses is found.
For momenta less than $\rm{500~MeV/c}$ an in-medium $\omega$ meson mass
of
$\rm{M_{medium}=[722_{-2}^{+2}(stat)_{-5}^{+35}(syst)]}~\rm{MeV/c^2}$
has been deduced at an estimated average nuclear
density of $0.6~\rho_0$.

\section{Summary}

The systematic study of total cross sections for single $\pi^\circ$,
$\eta$, and $\pi\pi$ production over a series of nuclei has not
provided an obvious hint for a depletion of resonance yield.
The observed reduction and change of shape in the second resonance region
are mostly as expected from absorption effects, Fermi smearing and
Pauli blocking, and collisional broadening.
The sum of experimental meson cross sections
for neutral and mixed charged states
between 400 and 800~MeV demonstrates the persistence
of the second resonance bump when at least one neutral meson is observed.
It has to be concluded that the medium modifications leading to the depletion of
cross section in nuclear photoabsorption are a subtle interplay of effects.
Their investigation and the rigorous comparison to theoretical models requires
the detailed study of differential cross sections and a deeper understanding of
meson production in the nuclear medium.
The recent experimental investigation of $\omega$ photoproduction from nuclei
has been presented as one detailed study of medium modification of hadrons.
First evidence for a lowering of the $\omega$ mass in the nuclear medium has
been observed.

\end{document}